\documentclass[conference]{IEEEtran}
\IEEEoverridecommandlockouts
\usepackage{cite}
\usepackage{amsmath,amssymb,amsfonts}
\usepackage{algorithmic}
\usepackage{graphicx}
\usepackage{textcomp}
\usepackage{xcolor}
\def\BibTeX{{\rm B\kern-.05em{\sc i\kern-.025em b}\kern-.08em
    T\kern-.1667em\lower.7ex\hbox{E}\kern-.125emX}}
\begin{document}

\title{A Study on E2E Performance Improvement of Platooning Using Outdoor LiFi\\
\thanks{This work was supported by JSPS KAKENHI Grant Number JP22H03585.}
}   

\author{
\IEEEauthorblockN{Zhiyi Zhu\textsuperscript{1,*}, 
 Eiji Takimoto\textsuperscript{2}, 
 Patrick Finnerty\textsuperscript{1},
 Chikara Ohta\textsuperscript{1}}
\\
\IEEEauthorblockA{\textsuperscript{1}\textit{Graduate School of System Informatics, Kobe University, Kobe, Japan}}
\IEEEauthorblockA{\textsuperscript{2}\textit{Information Technology Center, Nara Women's University, Nara, Japan}}
\IEEEauthorblockA{* Corresponding author email: syu@port.kobe-u.ac.jp}
}

\maketitle

\begin{abstract}
Platooning within autonomous vehicles has proven effective in addressing driver shortages and reducing fuel consumption.
However, as platooning lengths increase, traditional C-V2X (cellular vehicle-to-everything) architectures are susceptible to end-to-end (E2E) latency increases. 
This is due to the necessity of relaying information through multiple hops from the leader vehicle to the last vehicle. 
To address this problem, this paper proposes a hybrid communication architecture based on a simple simulation that integrates light fidelity (LiFi) and C-V2X. 
The proposed architecture introduces multiple-leader vehicles equipped with outdoor LiFi communication nodes in platoons to achieve high-speed and low-delay communication between leader vehicles, which reduces E2E delay.
\end{abstract}


\section{Introduction}
\begin{figure}
    \centering
    \includegraphics[width=0.9\linewidth]{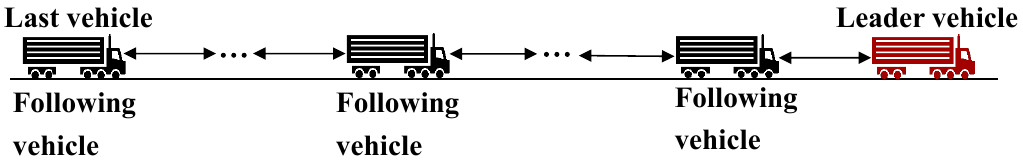}
    \caption{Platooning}
    \label{fig:common_platoon}
\end{figure}

Platooning is a critical application of connected autonomous vehicles (CAVs) that has garnered significant attention due to its notable advantages in enhancing traffic efficiency and reducing fuel consumption \cite{10077432}, as shown in Fig. \ref{fig:common_platoon}. 
%The stability and safety of platooning, particularly the implementation of cooperative adaptive cruise control (CACC) functionality, heavily rely on robust and low-latency vehicle-to-everything (V2X) communication \cite{ZHANG2023508}. 
However, traditional communication methods face challenges in long vehicle platoons. 
Information is typically transmitted via multi-hop relaying from the leader vehicle to the following vehicles. 
As the queue length increases, these delays by relaying will accumulate \cite{10683134}. 
For the rear vehicles, the end-to-end (E2E) performance may deteriorate, increasing the safety risks of the queue and potentially limiting its actual length.

Visible light communication (VLC) has recently begun to be applied in the field of vehicle-to-vehicle (V2V) communication. 
Early research has developed VLC prototypes, achieving low-latency communication of 36 milliseconds within a 30-meter range and successfully applying it to a simulated queue consisting of three vehicles \cite{7535434}.
Recently, LiFi technology, which is a VLC solution, has attracted significant attention. 
Researchers are gradually exploring its potential applications in vehicle-to-vehicle communication. 
The primary benefit of LiFi stems from its exploitation of the spectrum-rich visible and infrared light spectra, thereby circumventing the congestion and interference that plague the RF frequency band \cite{10683127}.
Researchers developed a lightweight V2V communication and perception integration system using vehicle headlights and taillights for information exchange and state perception between adjacent vehicles, with an effective communication range of approximately 5 meters \cite{10683127}.
Another study utilized vehicle cameras for optical communication to verify the identity of convoy members within a distance of 20 to 60 meters, thereby enhancing system security \cite{10571161}. 
Additionally, experimental results have demonstrated that LiFi can achieve 4.8 Gbps point-to-point transmission at a distance of 500 meters outdoors \cite{10529519}.

However, these studies primarily focus on short-range communication and have not explored how to utilize LiFi technology to address end-to-end (E2E) latency issues in long vehicle convoys. 
This is due to the inherent latency accumulation inherent to the traditional C-V2X multi-hop relay by sensing‑based semi‑persistent scheduling (SB‑SPS) \cite{10683134}. 

Therefore, this paper aims to address the E2E delay problem in long vehicle platoons. 
We propose a hybrid communication architecture that integrates LiFi and C-V2V (Cellular V2V). 
Specifically, we refer to recent models that utilize laser light sources and telescope optical systems to achieve point-to-point data transmission over hundreds of meters in outdoor long-range LiFi communication. 
The deployment of such LiFi systems on multiple leader vehicles in a platoon enables direct high-speed communication between leader vehicles, avoiding the delay accumulation caused by multi-hop relaying within C-V2V.

\section{Methodology and Simulation}

\begin{figure}[t]
    \centering
    \includegraphics[width=0.9\linewidth]{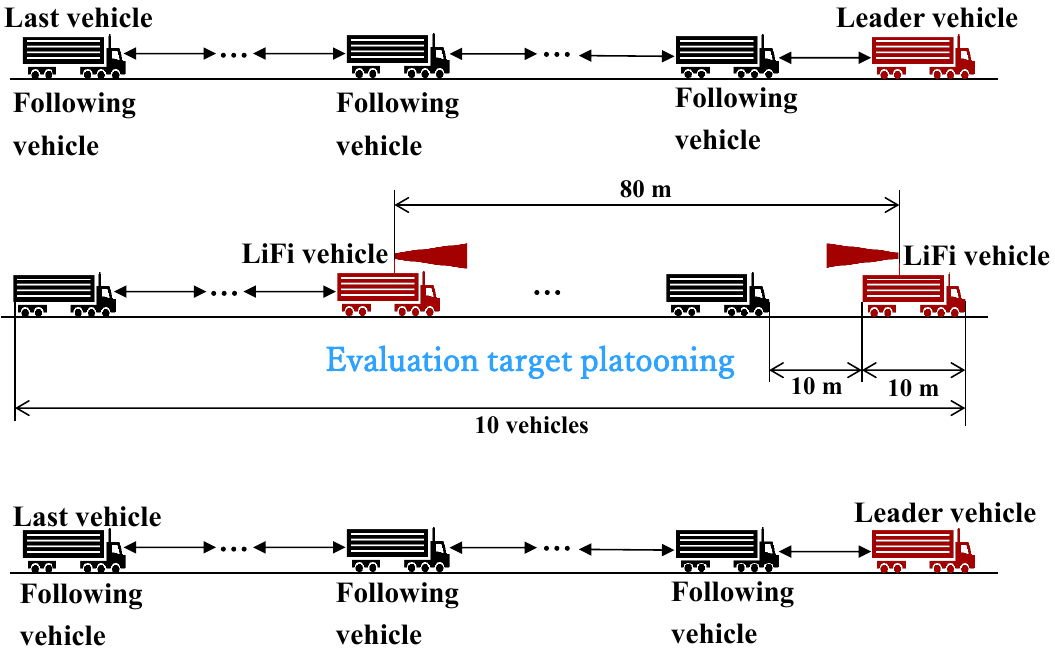}
    \caption{Simulation environment}
    \label{fig:sim-env}
\end{figure}

Three vehicle platoons are considered in the simulation environment, as shown in Fig. \ref{fig:sim-env}. 
The vehicle platoon in the middle is considered the target platoon for E2E delay evaluation, while the other two are interference platoons.
In the bidirectional leader (BDL) and predecessor‑leader following (PLF) information flow (IF) architecture, all three vehicle platoons use C-V2X communication to transmit data from the leader vehicle to the last vehicle. 
In the LiFi-based hybrid IF architecture, vehicles 1 and 6 use LiFi communication, while the remaining vehicles use C-V2X communication.

To evaluate E2E latency, we evaluate our LiFi-based hybrid IF architecture against two C‑V2V baselines: BDL and PLF IF architectures. 
In BDL, leader 1 sends data to vehicle 6 in one hop, after which vehicle 6 forwards it to vehicle 10 via multi‑hop. 
PLF relies entirely on multi‑hop relaying from leader 1 to vehicle 10. 
By contrast, our scheme uses a high‑capacity LiFi link from leader 1 to leader 6 (vehicle 6), which then distributes the message to vehicle 10.

\begin{figure}[t]
    \centering
    \includegraphics[width=0.65\linewidth]{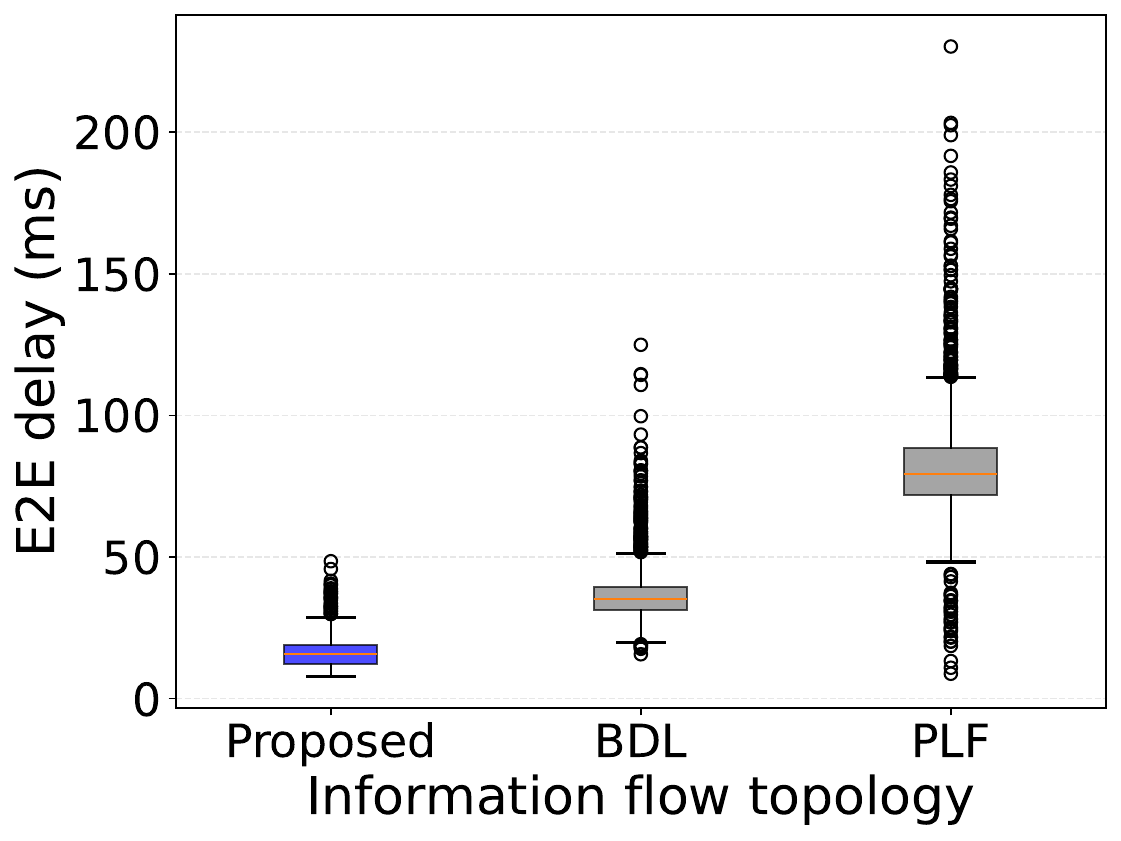}
    \caption{E2E delay between the first and last vehicles}
    \label{fig:e2e-res}
\end{figure}

We used the simulator provided by \cite{10077432}, and simulation parameters appear in Table \ref{tab:sim_params}, where $L_g$ is defined as the geometric diffusion loss, which is proportional to the square of the transmission distance; $d$ is the transmission distance, measured in kilometers; $L_{align}$ defined as the alignment loss, is contingent upon the angular deviation between the transmitter and receiver, other simulation parameters can be confirm in the Table.
Fig. \ref{fig:e2e-res} shows that the proposed LiFi-based hybrid communication architecture reduces E2E delay between the first and last vehicles relative to BDL and PLF, owing to the dual‑leader topology and the LiFi inter‑leader link.

\begin{table}[t]
\centering
\caption{Simulation Parameters}
\label{tab:sim_params}
\begin{tabular}{lc}
\hline
\textbf{Parameter} & \textbf{Value} \\
\hline
\multicolumn{2}{c}{5G V2X Communication Parameters} \\
\hline
Carrier frequency & 5.9 GHz \\
Number of platoons & 3 \\
Platoon size & 10 vehicles \\
Vehicle length & 10 m \\
Inter-vehicle distance & 10 m \\
Number of subchannels & 4 \\
Number of TTIs per period & 100 \\
Resource reservation interval (RRI) & 100 ms \\
Transmission power & 23 dBm \\
Sensing window & 1000 ms \\
Selection window & [1, 20] ms \\
Resource keeping probability & 0.4 \\
Reselection counter range & [5, 15] \\
Noise power & -114 dBm \\
Data packet size & 300 bytes \\
Packets per RRI & 10 \\
Path loss model & $PL = 40\log_{10}(d) + 11.82$ \\
\hline
\multicolumn{2}{c}{LiFi Communication Parameters} \\
\hline
Wavelength & 905 nm \\
Optical power & 0.45 W \\
Modulation bandwidth & 2.5 GHz \\
Receiver bandwidth & 1.4 GHz \\
Photodetector area & $10^{-4}$ m$^2$ \\
Photodetector responsivity & 0.5 A/W \\
Transmitter beam divergence & 5$^\circ$ \\
Receiver field of view & 30$^\circ$ \\
Noise current & $10^{-8}$ A \\
Ambient light current & $10^{-7}$ A \\
Atmospheric loss & 0.1 dB/km \\
Path loss & $L = L_g + 0.1d + L_{align}$ [dB] \\
\hline
\end{tabular}
\end{table}

\section{Conclusion}

In this paper, we propose a communication architecture combining LiFi and C-V2V based on outdoor long-range communication to address the issue of increased E2E delay in long vehicle queues. Through simulation, we validate the advantages of the proposed communication architecture over the C-V2V in terms of E2E delay performance. In future work, we will consider verifying the practicality of the proposed framework through experimental methods.

\bibliographystyle{IEEEtran}
\bibliography{ref}

\begin{thebibliography}{1}
\providecommand{\url}[1]{#1}
\csname url@samestyle\endcsname
\providecommand{\newblock}{\relax}
\providecommand{\bibinfo}[2]{#2}
\providecommand{\BIBentrySTDinterwordspacing}{\spaceskip=0pt\relax}
\providecommand{\BIBentryALTinterwordstretchfactor}{4}
\providecommand{\BIBentryALTinterwordspacing}{\spaceskip=\fontdimen2\font plus
\BIBentryALTinterwordstretchfactor\fontdimen3\font minus
  \fontdimen4\font\relax}
\providecommand{\BIBforeignlanguage}[2]{{%
\expandafter\ifx\csname l@#1\endcsname\relax
\typeout{** WARNING: IEEEtran.bst: No hyphenation pattern has been}%
\typeout{** loaded for the language `#1'. Using the pattern for}%
\typeout{** the default language instead.}%
\else
\language=\csname l@#1\endcsname
\fi
#2}}
\providecommand{\BIBdecl}{\relax}
\BIBdecl

\bibitem{10077432}
M.~Parvini, M.~R. Javan, N.~Mokari, B.~Abbasi, and E.~A. Jorswieck, ``Aoi-aware
  resource allocation for platoon-based c-v2x networks via multi-agent
  multi-task reinforcement learning,'' \emph{IEEE Transactions on Vehicular
  Technology}, vol.~72, no.~8, pp. 9880--9896, 2023.

\bibitem{10683134}
K.~Murata, K.~Sanada, H.~Hatano, and K.~Mori, ``An improved sb-sps scheme with
  low transmission delay for relay transmission in vehicle platooning with
  c-v2x sidelink communication,'' in \emph{2024 IEEE 99th Vehicular Technology
  Conference (VTC2024-Spring)}, 2024, pp. 1--5.

\bibitem{7535434}
M.~Y. Abualhoul, O.~Shagdar, and F.~Nashashibi, ``Visible light inter-vehicle
  communication for platooning of autonomous vehicles,'' in \emph{2016 IEEE
  Intelligent Vehicles Symposium (IV)}, 2016, pp. 508--513.

\bibitem{10683127}
Y.~Song, R.~Mo, P.~Zhang, C.~Wang, Z.~Sheng, Y.~Sun, and Y.~Yang,
  ``Vehicletalk: Lightweight v2v network enabled by optical wireless
  communication and sensing,'' in \emph{2024 IEEE 99th Vehicular Technology
  Conference (VTC2024-Spring)}, 2024, pp. 1--5.

\bibitem{10571161}
M.~Plattner and G.~Ostermayer, ``Vehicle-to-vehicle optical camera
  communications for platoon verification,'' in \emph{2024 IEEE Wireless
  Communications and Networking Conference (WCNC)}, 2024, pp. 01--06.

\bibitem{10529519}
C.~Chen, S.~Das, S.~Videv, A.~Sparks, S.~Babadi, A.~Krishnamoorthy, C.~Lee,
  D.~Grieder, K.~Hartnett, P.~Rudy, J.~Raring, M.~Najafi, V.~K. Papanikolaou,
  R.~Schober, and H.~Haas, ``100 gbps indoor access and 4.8 gbps outdoor
  point-to-point lifi transmission systems using laser-based light sources,''
  \emph{Journal of Lightwave Technology}, vol.~42, no.~12, pp. 4146--4157,
  2024.

\end{thebibliography}

\end{document}